# Mechanical magnetometry of Cobalt nanospheres deposited by focused electron beam at the tip of ultra-soft cantilevers

Hugo Lavenant[1], Vladimir Naletov[1,2], Olivier Klein[1], Grégoire de Loubens[1,*], Laura Casado[3], José María De Teresa[3,4]

1. Service de Physique de l'État Condensé (CNRS URA 2464), CEA Saclay, 91191 Gif-sur-Yvette, France
2. Institute of Physics, Kazan Federal University, Kazan 420008, Russian Federation
3. Laboratorio de Microscopías Avanzadas (LMA), Instituto de Nanociencia de Aragón (INA), Universidad de Zaragoza, Mariano Esquillor 50018 Zaragoza, Spain
4. Instituto de Ciencia de Materiales de Aragón (ICMA), Departamento de Física de la Materia Condensada, Universidad de Zaragoza-CSIC, Pedro Cerbuna 12, 50009 Zaragoza, Spain

*Corresponding author: gregoire.deloubens@cea.fr

**Abstract**

Using focused-electron-beam-induced deposition, Cobalt magnetic nanospheres with diameter ranging between 100 nm and 300 nm are grown at the tip of ultra-soft cantilevers. By monitoring the mechanical resonance frequency of the cantilever as a function of the applied magnetic field, the hysteresis curve of these individual nanospheres are measured. This enables to evaluate their saturation magnetization, found to be around 430 emu/cm$^3$ independently of the size of the particle, and to infer that the magnetic vortex state is the equilibrium configuration of these nanospheres at remanence.

**Keywords**

**Introduction**

The magnetic functionalization of micro-fabricated cantilevers is crucial for magnetic force microscopy (MFM), a widely used imaging tool in the field of nanomagnetism [1, 2], and for magnetic resonance force microscopy (MRFM), a technique which combines MFM and magnetic resonance imaging to investigate spin dynamics at the nanoscale [3, 4]. For quantitative analysis of the mechanical signal [5, 6], it is important to be able to carefully control and characterize the nanomagnet at the tip of the cantilever, whose actual size defines both the spatial resolution and the sensitivity. In MRFM, moreover, this figure of merit is governed by the very large magnetic field gradients produced in the proximity of the nanomagnet [7]. This can be much improved by attaching a magnetic nanoparticle instead of depositing a magnetic layer onto the tip. Here the quality of the magnet (size, shape, magnetization, coercivity, remanence) is of the utmost importance. But detecting tiny mechanical forces also commands the use of ultra-soft cantilevers [8] and it is not easy to incorporate high-quality nanomagnets to such mechanical oscillators, because conventional fabrication methods are not compatible with their extreme softness.

In state-of-the-art MRFM experiments, the control of the field gradient source dominates the technical constraints. It is thus the sample rather than the magnetic probe that is attached at the end of the cantilever, while the field gradient is produced by a permanent nanomagnet placed underneath [9]. For improved versatility, however, it is preferable to have the nanomagnet directly at the tip of the cantilever. A tremendous effort has been put in a nanofabrication process that enables an on-chip integration of Co nanomagnets with ultra-soft cantilevers [10]. It is also possible either to glue a micron-size spherical magnetic probe at the end of a cantilever [11], or to attach a tiny permanent magnet and to shape it by focused ion beam [12], or even to use an iron filled carbon nanotube [13]. In all these cases, having the very end of the probe (which has to be approached in the close vicinity of the sample) with good magnetic properties is a challenge. We also note that in the case of MRFM applied to ferromagnetic nanostructures, the optimum sensitivity, or filling factor, requires a specific size for the nanomagnet, which should be of the same order than the studied sample [11]. There are, however, few methods for the synthesis of nanomagnets in the 10 nm – 300 nm range.

In that context, the possibility to grow high-quality cobalt nanoparticles by focused-electron-beam-induced deposition (FEBID) [14-16] opens an interesting new route to attaching *in-situ* nanosize magnets at the tip of cantilevers. In fact, no micromanipulation is required in that case to position the nanomagnet at the apex of the cantilever beam. Such sub-micronic Co nanomagnets of roughly hemispherical shape have been recently used in some MRFM experiments [17, 18]. Still, more effort has to be put in this technology in order to be applied to ultra-sensitive, quantitative MFM or MRFM studies. Firstly, one needs to control the geometry of nanomagnets grown by FEBID on cantilevers having very small spring constant (k < 0.01 N/m), for which vibrations during the deposition process might be an issue. Secondly, a detailed magnetic characterization has to be performed on these nanomagnets to check their quality and to investigate their magnetic configuration, which might be non trivial depending on the applied field. For instance, magnetic nanospheres are expected to have highly non-uniform equilibrium states at low field, and among them, topological singularities such as a vortex [19], a Bloch point [20], or a skyrmion [21].

In this work, we use FEBID to grow Co nanospheres with diameters ranging from 300 nm down to 100 nm at the tip of ultra-soft cantilevers, and cantilever magnetometry to characterize their magnetic properties. We first explain the nanofabrication method and the principles of this specific magnetization measurement. We then present the magnetization data obtained on the individual magnetic nanospheres and analyze them to extract their magnetic moment and saturation magnetization. Finally, we discuss their magnetic configuration at remanence and some future possible work.

**Methods**

Using the Dual Beam facility of the LMA at Universidad de Zaragoza, we have grown by FEBID Co nanospheres of nominal radius 300, 200 and 100 nm at the end of several cantilevers. Scanning electron microscopy (SEM) images of such nanospheres are presented in Figure 1. In this work, we have used soft commercial Olympus Biolevers in silicon nitride (nominal spring constant k = 6 mN/m, resonance frequency $f_c$ = 13 kHz, quality factor 2000 < Q < 4000 under vacuum), which are well adapted to MRFM studies [6, 11, 22–26] due to their excellent force sensitivity, $F_{min}$ ~ 0.7 fN/√Hz at room temperature. These cantilevers are also convenient to image in a SEM thanks to their thin gold coating. By delicately placing the cantilevers beam on a support, it was possible to prevent their motion during the deposition process. We were then able to grow by FEBID roughly spherical Co nanoparticles having the requested lateral dimensions at the very end of the special V-shape tip of the cantilever. The precursor used for growing these Cobalt nanospheres was $Co_2(CO)_8$ as previous work has demonstrated the growth of ultra-small magnetic structures (< 30 nm) using this approach [27, 28]. When the precursor was introduced close to the cantilever tip, the chamber vacuum pressure changed from $1 \cdot 10^{-6}$ mbar (base pressure) to $8.5 \cdot 10^{-6}$ mbar (process pressure). Co nanospheres of nominal radius 300 and 200 nm were grown at 5 kV and 50 pA using the high-resolution (in-lens) mode II. Co nanospheres of nominal radius 100 nm were grown at 5 kV and 25 pA. Using these growth conditions, the Co purity reached was 75±5% at., as measured by EDX. Previous work has shown that the microstructure of these cobalt deposits consists of polycrystalline cobalt grains inside a carbonaceous amorphous matrix [29]. Due to the polycrystalline nature of the deposits, their magnetic anisotropy is expected to be governed by shape anisotropy [30].

After the growth of Co nanoparticles, each cantilever is kept under static vacuum during a few days. It is then introduced for characterization in the vacuum chamber (P < $10^{-5}$ mbar) of an MRFM microscope sitting in between the poles of an electromagnet and operated at a stabilized temperature of 290 K [11]. A standard laser deflection technique is used to monitor the displacement of the cantilever, whose mechanical characteristics are determined from noise measurements [8]. Using a piezoelectric bimorph and a feedback electronic circuit based on a phase lock loop, we can also track its resonance frequency while maintaining its vibration amplitude constant (in this work, typically 10 nm, corresponding to roughly 30 times the Brownian motion amplitude of the cantilever).

In magnetometry measurements, the mechanical resonance frequency of the cantilever is monitored as a function of the applied magnetic field. If the individual nanomagnet attached on the cantilever has some shape or crystalline magnetic anisotropy, the measured frequency shift *vs.* the spatially homogeneous magnetic field originates from the magnetic torque acting on the cantilever [13, 31–33]. To perform magnetometry measurements of a nanomagnet in which no anisotropy is expected (*e.g.*, an amorphous magnetic nanosphere), we plunge the tip of the cantilever in the field gradient produced by a magnetic cylinder, as indicated in the experimental sketch of Figure 2. In this case, the effective spring constant of the cantilever depends on the magnetic force acting on it, which is proportional to both the nanomagnet's magnetic moment *m* and to the field gradient $dB_z/dz$, considered to be along z. The resulting cantilever frequency shift due to the presence of the magnetic moment at its end then writes:

$$\frac{\Delta f_c}{f_c} = -\frac{m}{2k}\left(\frac{\partial^2 B_z}{\partial z^2}\right)_{|z_0} \qquad \text{(Equation 1)}$$

where $z_0$ indicates the equilibrium position of the nanomagnet in the field gradient [34]. Hence, if k and $d^2B_z/dz^2$ are precisely known in Equation 1, a quantitative determination of *m* is possible.

In our experimental setup, the source of field gradient is a millimeter long cylinder of $Co_{64}Fe_{6.5}Ni_{1.5}Si_{14}B_{14}$ alloy having a magnetization saturation of 510 emu/cm$^3$ determined by SQUID magnetometry [35]. Its diameter determined by SEM imaging is approximately 16 μm and it is surrounded by a 4 μm thick glass sheath to protect it against oxidation (see inset of Figure 2). It was chosen because thanks to its shape anisotropy, it is expected to be fully saturated along its symmetry axis z even at low applied magnetic field. Assuming a perfect cylindrical shape, it is also possible to calculate analytically the magnetic stray field and field gradients above it [36]. For instance, at a distance of 4 μm on the z axis above our cylinder, we obtain that $d^2B_z/dz^2$ can reach up to $4.3 \cdot 10^9$ G/cm$^2$. In such a large field gradient, a magnetic moment $m = 10^{-13}$ emu (= $10^7$ Bohr magnetons) would produce a detectable frequency shift of 0.5 Hz, corresponding to more than 10% of the full line width of the mechanical resonance of the cantilever.

**Results**

In order to calibrate our cantilever magnetometry experiment, we have first studied a well characterized nanomagnet attached at the end of a Biolever. A SEM image of it is presented in the inset of Figure 3a. It is a nanosphere of diameter 700 nm made of an amorphous FeSi alloy with 3% in mass of silicon, which was already employed in several MRFM experiments [6, 23–25]. In these studies, the magnetic moment of this MRFM probe, $m = (2.5 \pm 0.5) \cdot 10^{-10}$ emu, was inferred from its stray field, that can be calculated by assuming a punctual magnetic moment at the center of the sphere. Experimentally, one can indeed readily measure the field shift of the ferromagnetic resonance of the sample due to the stray field of the magnetic probe, placed at a known distance above it [11].

After positioning this 700 nm FeSi reference sphere at a distance of $13 \pm 2$ μm above the center of the magnetic cylinder, the cantilever frequency shift was recorded as a function of the magnetic field applied along the axis of the cylinder, see Figure 3a. Using the maximum experimental frequency shift of −290 Hz measured at large applied field with respect to zero field and the estimation of the field gradient $d^2B_z/dz^2 = 10^9$ G/cm$^2$ at the position where the measurement is performed, Equation 1 yields a magnetic moment $m = 3 \cdot 10^{-10}$ emu, in good agreement with the expected value. Since the magnetization of the cylinder is always aligned along the applied field, parallel to its axis, the raw data of cantilever frequency *vs.* field can easily be translated into a magnetization curve. The fact that the obtained curve in Figure 3b does not exhibit any sizable hysteresis and is linear in field below saturation indicates that the reference probe has a very weak anisotropy, as expected for an amorphous magnetic sphere. Its saturation field $H_s$ of about 6.5 kOe also makes sense. For a perfect sphere without crystalline anisotropy, it is indeed only governed by demagnetizing effects, $H_s = 4\pi M_s/3$. The saturation magnetization of the reference FeSi sphere is thus $M_s = 1550$ emu/cm$^3$, which, multiplied by its volume $V = (4/3)\pi R^3$, where R is the radius of the sphere, leads to a magnetic moment of $2.8 \cdot 10^{-10}$ emu. Hence, the previously reported MRFM studies and informations that can be extracted from the magnetometry data match well together, and we have indicated in Table 1 the final values of the radius R, magnetic moment *m*, and saturation magnetization $M_s$ of the reference FeSi probe with the experimental error bars.

We now continue with the cantilever magnetometry measurements of the FEBID grown Co nanospheres. In Figure 4, we compare the relative frequency shifts measured as a function of field for the reference FeSi probe, a 300 nm Co nanosphere, and a 100 nm Co nanosphere. These three data sets have been obtained at a distance of $11 \pm 2$ μm above the magnetic cylinder. The maximum relative frequency shift measured for the reference probe at this position is $\Delta f_c/f_c = 4.1\%$. It is $\Delta f_c/f_c = 0.067\%$ for the 300 nm Co particle, and $\Delta f_c/f_c = 0.0046\%$ for the 100 nm Co particle. Since the field gradient in which the different nanomagnets are plunged is approximately the same, one can infer that the magnetization of the 100 nm Co sphere is about 14.5 times smaller than the one of the 300 nm Co sphere, itself about 61 times smaller than the 700 nm FeSi reference sphere. In order to get more accuracy on the determination of the magnetic moments of the Co nanospheres, we have repeated these measurements at various tip-cylinder separations, ranging from 4 to 15 μm. We extract that on average, $(\Delta f_c/f_c)_{[700 \text{ nm FeSi}]} = (60 \pm 10) (\Delta f_c/f_c)_{[300 \text{ nm Co}]} = (850 \pm 50) (\Delta f_c/f_c)_{[100 \text{ nm Co}]}$. Knowing the magnetic moment of the reference probe, those of the Co nanospheres are extracted and reported in Table 1.

| Nanosphere | Radius (nm) | Magnetic moment (emu) | Magnetization (emu/cm$^3$) |
|---|---|---|---|
| 700 nm FeSi reference | $350 \pm 30$ | $(2.8 \pm 0.7) \cdot 10^{-10}$ | $1550 \pm 70$ |
| 300 nm FEBID-Co | $145 \pm 4$ | $(4.9 \pm 1) \cdot 10^{-12}$ | $430 \pm 80$ |
| 100 nm FEBID-Co | $57 \pm 6$ | $(3.2 \pm 1) \cdot 10^{-13}$ | $430 \pm 80$ |

**Table 1.** Summary of cantilever magnetometry measurements performed on three different magnetic nanospheres.

In Figures 5 and 6, we present the hysteresis curves of the 100 nm and 300 nm Co nanospheres, respectively. SEM images, useful to check the shape of these nanomagnets, are presented in the insets of Figures 5a and 6a. In order to improve the signal to noise ratio, these measurements have been performed at smaller tip-cylinder separation (< 4 µm), in regions of stronger field gradients above the magnetic cylinder, resulting in similar maximal relative frequency shifts of about $\Delta f_c/f_c$ = 0.25% for the 100 nm and 300 nm particles, see Figures 5a and 6a. The translation of these data into magnetization curves are presented in Figures 5b and 6b. At first, these magnetization curves look quite similar to that of the FeSi reference sphere. They both exhibit a linear variation at low field, and a saturation at larger field, with very weak hysteresis. The most striking difference with the reference probe is the value of the saturation field, estimated to be 0.7 ± 0.2 kOe. But in these measurements, the additional stray field from the cylinder cannot be neglected, since it could be as large as 2 kOe at small distances above the cylinder. In order to get a quantitative estimation of the saturation field, one should rather look at the data of Figure 4, where the tip-cylinder separation is much larger and the additional stray field from the cylinder is only a few hundreds of Oersteds. From these data, one would estimate that the saturation field of the Co nanospheres is $H_s$ = 1.8 ± 0.3 kOe, *i.e.*, a saturation magnetization of 430 ± 80 emu/cm$^3$. This value compares favorably with the one estimated from the ratio of the magnetic moment to the volume of the particles, found to be approximately 400 emu/cm$^3$.

**Discussion**

An interesting aspect of these magnetic nanospheres is that they exhibit nearly zero magnetization at remanence. Preliminary micromagnetic simulations of our 100 and 300 nm Co nanospheres, in which the magnetic parameters determined experimentally have been used, show indeed that in both cases, a magnetic vortex is nucleated at sufficiently low field at the center of the sphere. This flux closing configuration arises from the competition between exchange and dipolar interactions. By looking carefully at the hysteresis loop of the 300 nm particle (see inset of Figure 6b, that shows a zoom of the data at low field), one can notice that it is clearly non reversible, and that small magnetization jumps occur at well defined fields of ± 0.1 and ± 0.4 kOe, which might be the signature of vortex core nucleation, reversal, and annihilation, as in magnetic nanodisks [23, 37]. The fact that the hysteresis loop of the nanosphere of diameter 100 nm looks more squared that the one of diameter 300 nm is also an interesting experimental observation, which points towards the importance of finite size effects in the magnetization process of these nanomagnets.

To investigate further this point, one should perform thorough magnetometry measurements of nanospheres as a function of their diameter and a detailed simulation work. It would also be very useful to have and experimental access to the micromagnetic configuration of our nanospheres. For instance, electron holography microscopy would be well adapted to this task, as it is able to produce 2D maps of the magnetic induction inside and around a ferromagnetic material with very high spatial resolution, typically down to 5 nm [38, 39]. One could also try to map the magnetic field of the expected magnetic vortex core at the center of the spheres with scanning NV center magnetometry [40]. We would also like to stress that if the magnetic relaxation of the FEBID Co nanospheres is not too strong, their characteristic high frequency magnetization dynamics [19] could be investigated by MRFM [23].

The Co content of the nanospheres is thought to be the cause of the relatively low saturation magnetization of the nanospheres, about three times less than the one of bulk Cobalt. It might be improved with further growth optimization. In fact, it was shown that 2D FEBID Co nanostructures of high purity (>95% at.) can be grown in optimal conditions [15, 41]. In the case of 3D Co structures, the Co content is typically found to be around 85% [42]. Thus, we expect that fine tuning of the growth conditions could increase the Co content currently obtained in the nanospheres (75±5%), thus enhancing their magnetization. Such growth optimization would involve using lower electron beam current, exploring the influence of the dwell time and the refresh time, modifications of the scan strategy and minimization of the vibration of the tip during the fabrication.

We also point out that other materials different from Co, such as Fe, can be efficiently grown by FEBID [43, 44]. Other 3D geometries different from nanospheres might be useful to tailor the magnetic properties of the nanomagnets grown on cantilevers [42]. This would open a wide field for improving the spatial resolution and the sensitivity of magnetic force microscopy (MFM), the scanning probe method at the basis of MRFM.

**Conclusion**

As shown by sensitive cantilever magnetometry measurements, the Co nanospheres grown in this work by FEBID at the tip of ultra-soft cantilevers are ferromagnetic, and as such, could be very useful probes for MRFM [17, 18]. The small diameter of the grown nanospheres has great potential for achieving sub-100 nm resolution in future MRFM experiments. Further improvements during the growth could lead to Co nanospheres with higher magnetization, which would optimize the signal-to-noise ratio in MRFM experiments.

**Acknowledgements**

The research leading to these results has received funding from the European Union Seventh Framework Programme under Grant Agreement 312483 – ESTEEM2 (Integrated Infrastructure Initiative–I3), from the French Agence Nationale de la Recherche under Grant Agreement ANR-2010-JCJC-0410-01 – MARVEL, from Spanish Ministry of Economy and Competitivity through project No. MAT2011- 27553-C02, including FEDER funds, and by the Aragón Regional Government. Experimental help by Dr. Soraya Sangiao and Dr. J. M. Michalik, as well as the kind assistance of Dr. R. Lassalle-Ballier for micromagnetic simulations are warmly acknowledged.

**Figures**

**Abstract figure.** SEM image of a 200 nm Co nanosphere grown at the tip of an ultra-soft cantilever by focus electron beam induced deposition.

**Figure 1.** SEM images of Co nanospheres with diameter ranging from 300 nm down to 100 nm grown at the tip of an ultra-soft cantilever by focus electron beam induced deposition. Top images are front views of the tip, bottom images are side views.

**Figure 2.** Sketch of the experimental setup. The cantilever with the magnetic nanosphere at its tip is plunged in the field gradient of a magnetic cylinder (see SEM image in lower left inset).

**Figure 3.** Cantilever magnetometry of the 700 nm FeSi reference sphere. (a) Raw data of the cantilever frequency as a function of the applied field. The inset shows a SEM image of the measured magnetic nanosphere at the tip of the cantilever. (b) Corresponding magnetization curve.

**Figure 4.** Comparison of the relative frequency shifts of the cantilevers with the 700 nm FeSi reference sphere, the 300 nm Co nanosphere, and the 100 nm Co nanosphere as a function of the applied magnetic field. In these measurements, the separation between the source of field gradient (magnetic cylinder, see Fig.2) and the tip of the cantilevers is set to $11 \pm 2$ μm.

**Figure 5.** Cantilever magnetometry of the 100 nm Co nanosphere. (a) Raw data of the cantilever frequency as a function of the applied magnetic field. The inset shows a SEM image of the measured magnetic nanosphere at the tip of the cantilever. (b) Corresponding magnetization curve.

**Figure 6.** Cantilever magnetometry of the 300 nm Co nanosphere. (a) Raw data of the cantilever frequency as a function of the applied magnetic field. The inset shows a SEM image of the measured magnetic nanosphere at the tip of the cantilever. (b) Corresponding magnetization curve. The inset is a zoom of the behavior at low field, where characteristic jumps of the magnetization have been marked by arrows.

**Abstract figure**

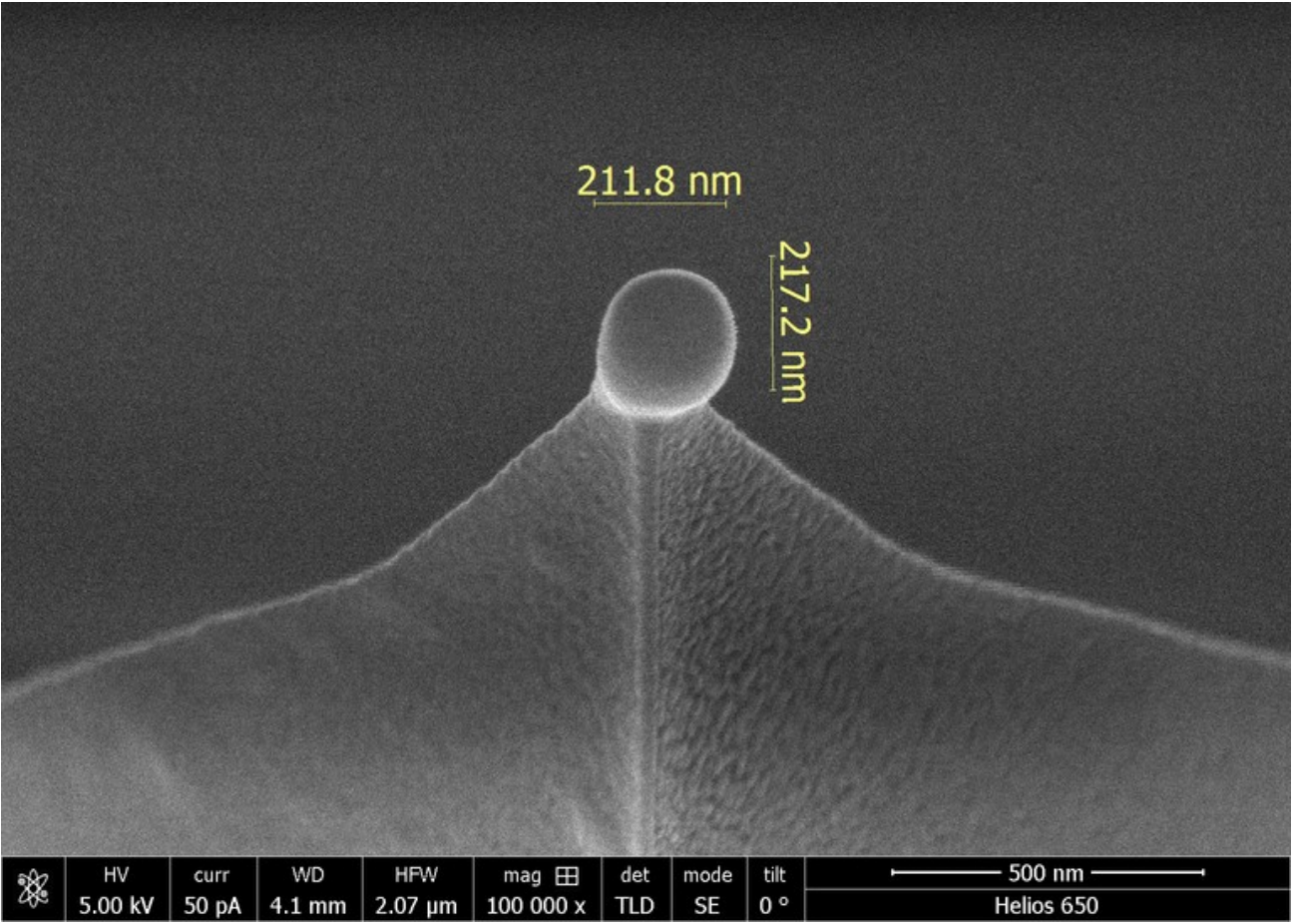

**Figure 1**

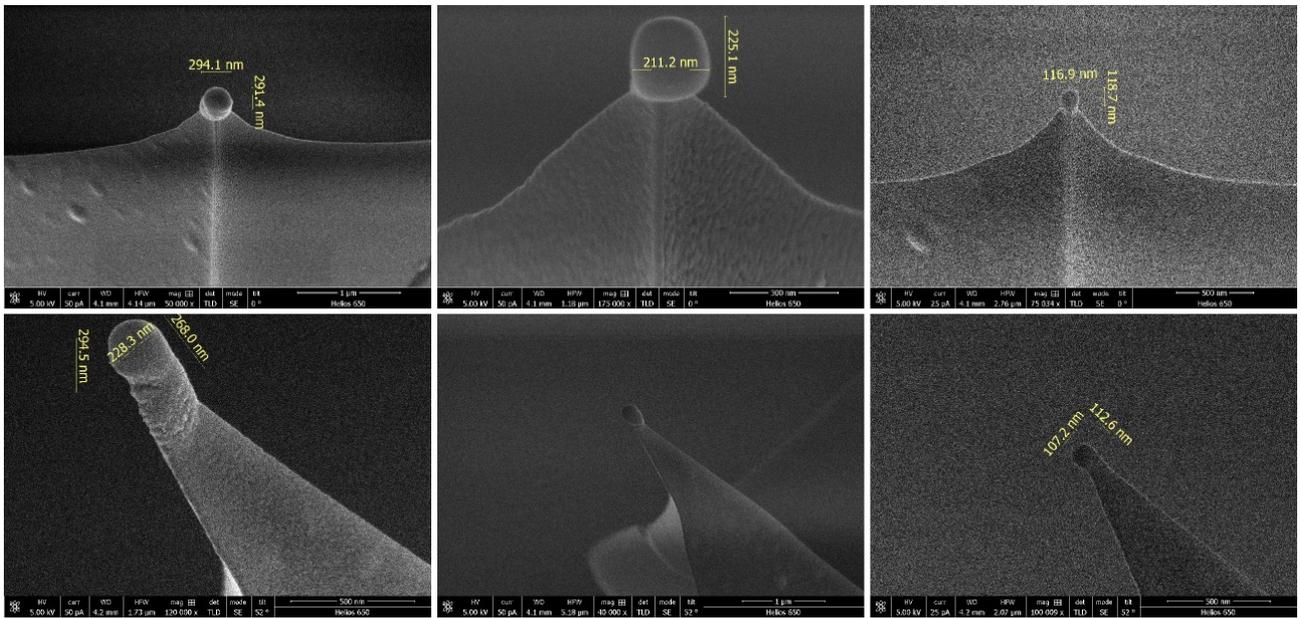

**Figure 2**

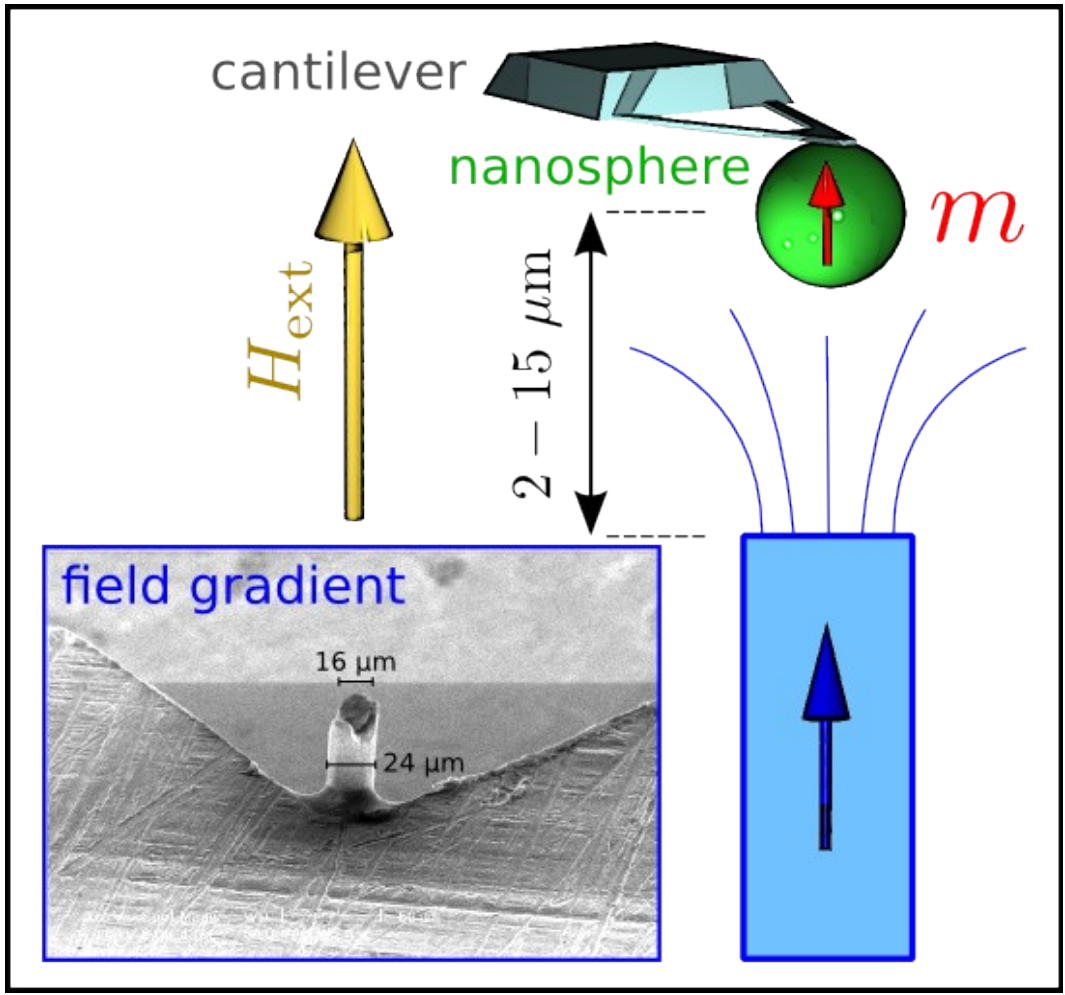

**Figure 3**

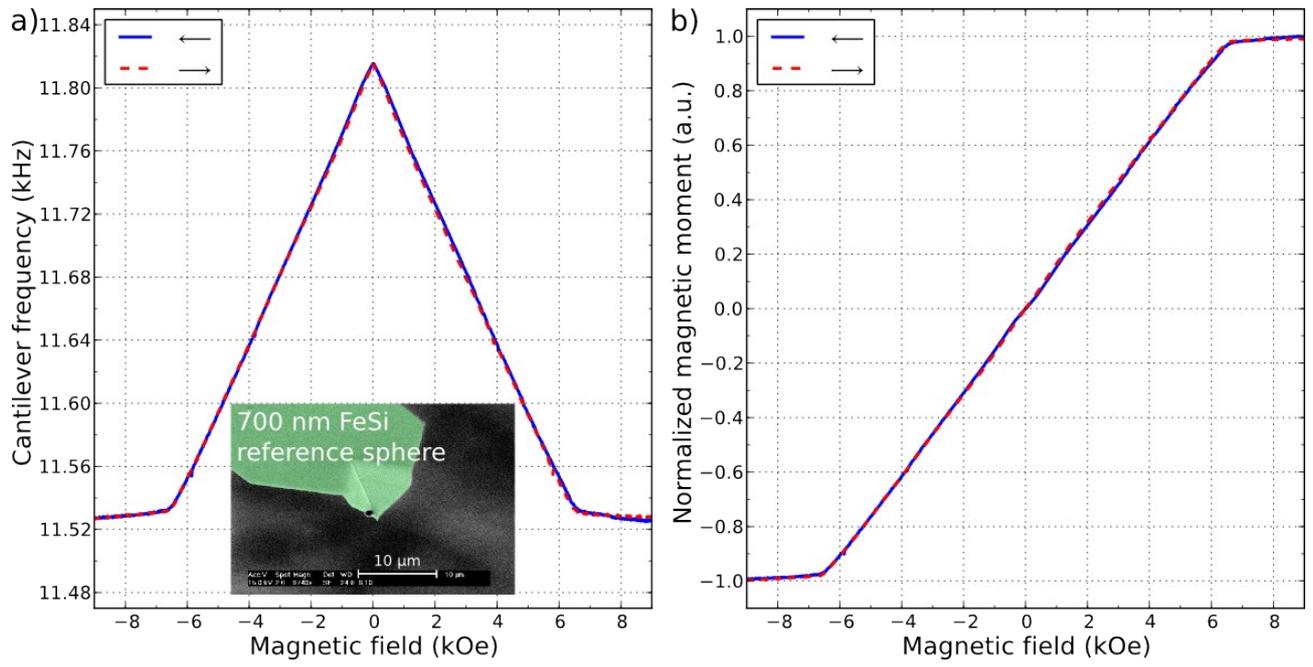

**Figure 4**

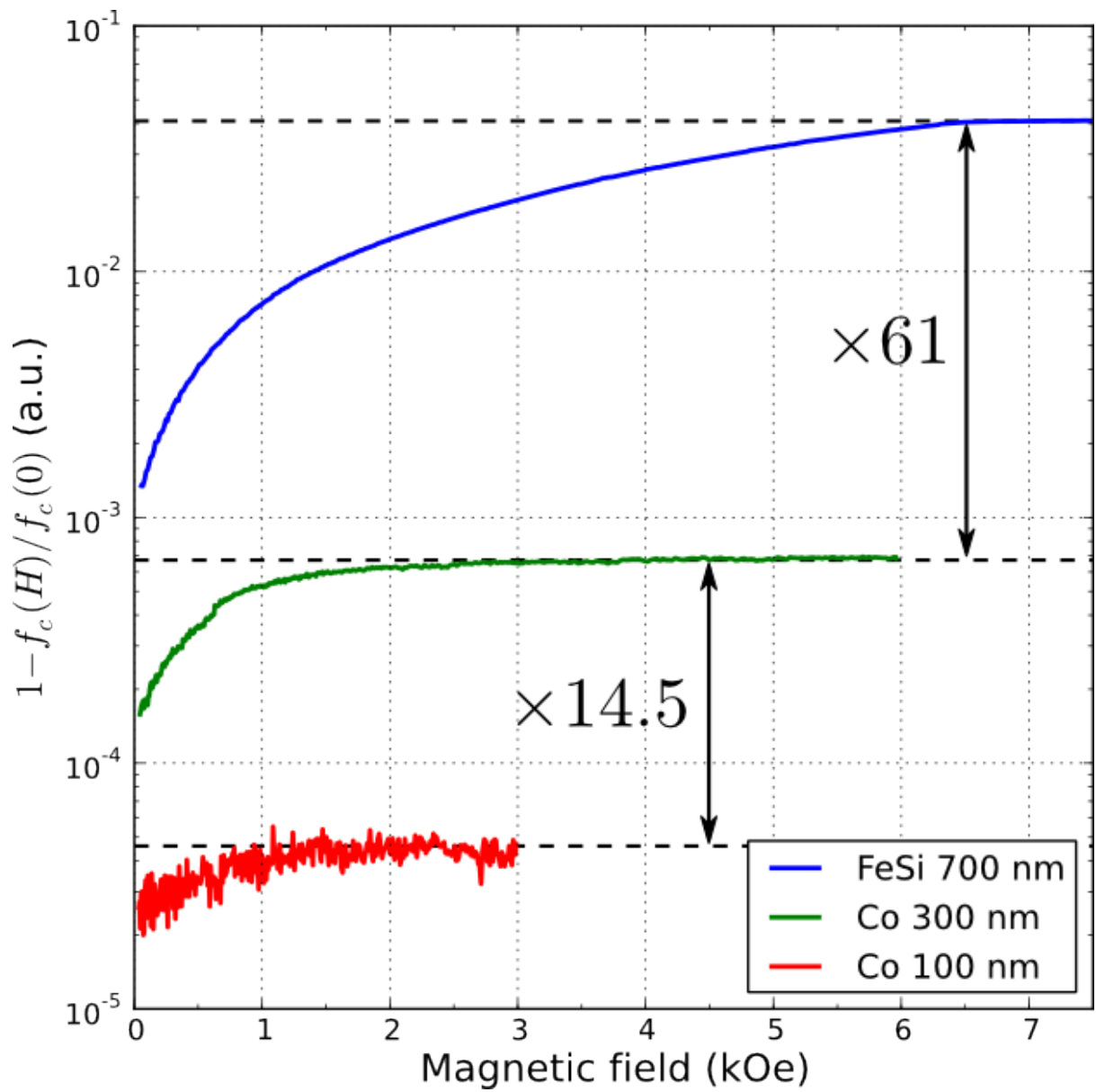

**Figure 5**

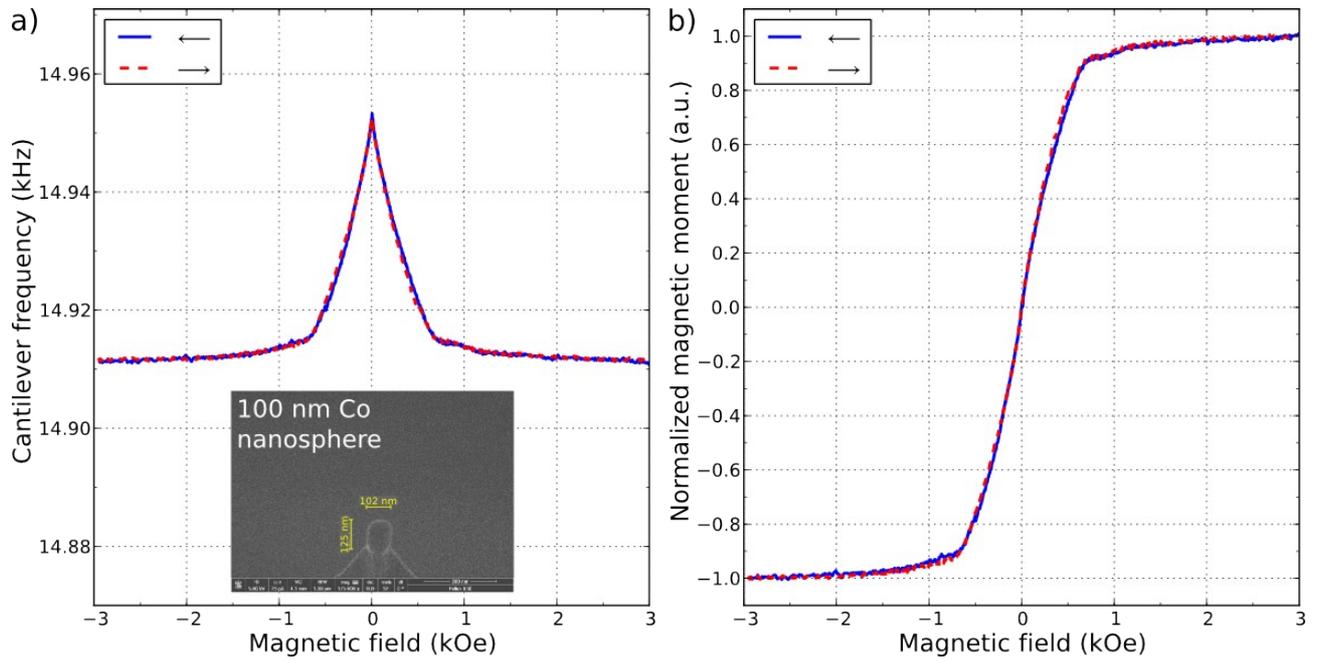

**Figure 6**

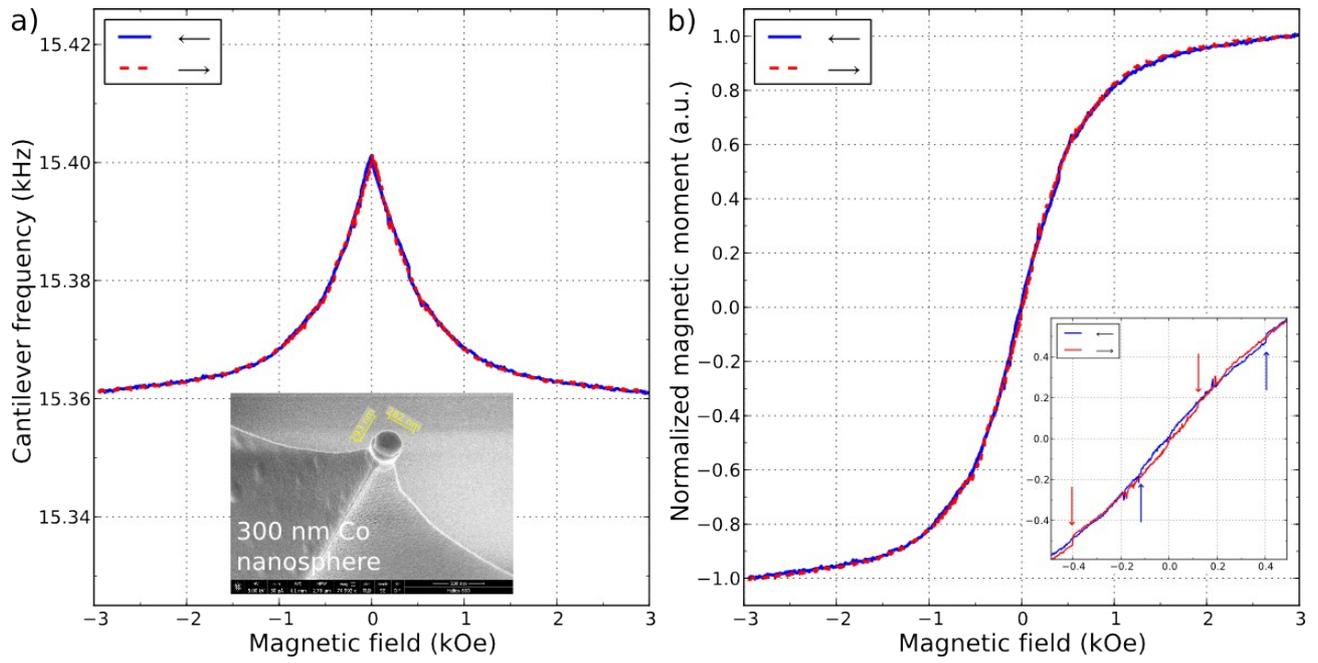